\documentclass[11.5pt,letterpaper]{article}
\usepackage{amssymb}
\usepackage{amsmath}
\usepackage{physics}
\usepackage{graphicx}
\usepackage{float}
\usepackage{slashed}
\usepackage{caption}
\usepackage{subcaption}
\usepackage{xcolor}
\usepackage{ulem}

\usepackage{multirow}
\usepackage{cite,color,xcolor,url}
\def\linkcolor{cyan!70!black}

\usepackage[
colorlinks=true
,urlcolor=\linkcolor
,anchorcolor=\linkcolor
,citecolor=\linkcolor
,filecolor=\linkcolor
,linkcolor=\linkcolor
,menucolor=\linkcolor
,linktocpage=true
,pdfproducer=medialab
,pdfa=true
]{hyperref}

\newcommand{\prth}[1]{\left(#1 \right)}

\newcommand{\LHC}{\textrm{LHC}}
\newcommand{\HLLHC}{\textrm{HL-LHC}}
\newcommand{\current}{\textrm{current}}
\newcommand{\SM}{\textrm{SM}}
\newcommand{\bins}{\textrm{bins}}

\newcommand{\mM}{\mathcal{M}}

\newcommand{\mn}{{\mu\nu}}

\title{
\begin{flushright}
{\normalsize IFT-UAM/CSIC-24-4}
\end{flushright}

\vspace*{1cm}

\sc Drell-Yan Bounds on Gapped Continuum Spectra
}

\author{
\textsc{E. Arganda}$^a$\footnote{{\tt \href{mailto:ernesto.arganda@uam.es}{ernesto.arganda@uam.es}}}~,\, \textsc{A. Delgado}$^b$\footnote{{\tt \href{mailto:adelgado2@nd.edu}{adelgado2@nd.edu}}}~,\, \textsc{A. Martin}$^b$\footnote{{\tt \href{mailto:amarti41@nd.edu}{amarti41@nd.edu}}}~, \textsc{E. Meg\'{\i}as}$^c$\footnote{{\tt \href{mailto:emegias@ugr.es}{emegias@ugr.es}}}~, \\ \\ \textsc{R. Morales}$^d$\footnote{{\tt \href{mailto:roberto.morales@fisica.unlp.edu.ar}{roberto.morales@fisica.unlp.edu.ar}}}~,\, \textsc{M. Quir\'os}$^e$\footnote{{\tt \href{mailto:quiros@ifae.es}{quiros@ifae.es}}}~,\, \textsc{T. Saxton}$^b$\footnote{{\tt \href{mailto:tsaxton@nd.edu}{tsaxton@nd.edu}}}
}

\begin{document}

\maketitle

\noindent $^a$ \textit{\large Departamento de Física Teórica and Instituto de F\'{\i}sica Te\'orica UAM-CSIC, Universidad Autónoma de Madrid, Cantoblanco, 28049 Madrid, Spain}

\noindent $^b$ \textit{\large Department of Physics, University of Notre Dame, 225 Nieuwland Hall
Notre Dame, IN 46556, USA}

\noindent $^c$ \textit{\large Departamento de F\'{\i}sica At\'omica, Molecular y Nuclear and Instituto Carlos I de F\'{\i}sica Te\'orica y Computacional,
Universidad de Granada, Avenida de Fuente Nueva s/n, 18071 Granada, Spain}

\noindent $^d$ \textit{\large IFLP, CONICET - Dpto. de Fisica, Universidad Nacional de La Plata,
C.C. 67, 1900 La Plata, Argentina}

\noindent $^e$ \textit{\large Institut de Fisica d'Altes Energies (IFAE) and BIST, Campus UAB
08193, Bellaterra, Barcelona, Spain}
\begin{abstract}
\large
Theories with gapped continuum spectra have gotten some attention, either as pure 4D models like unparticles, or in 5D realizations as certain soft walls constructions. In this paper, we derive experimental bounds from Drell-Yan processes ($pp \to \ell^+\ell^-$, $pp \to \ell^\pm \nu$) in a particular scenario where the electroweak bosons propagate in an extra dimension that produces a propagator with a continuum spectrum, on top of the isolated corresponding Standard Model pole. Using current LHC data we put a lower bound on the gap of 4.2 TeV (expected), 6.2 TeV (observed, bins with $< 10$ events combined) at 95\% CL, with some dependence in the observed limit on how low statistics bins are treated. We also study the limits for HL-LHC.
\end{abstract}
\vspace{1cm}

\newpage
\large
\section{Introduction}

The Standard Model (SM) of electroweak and strong interactions is firmly established as a valid theory for scales $Q\lesssim $~TeV by past and present high-energy and low-energy colliders. Still, it is well accepted that the model requires an ultraviolet (UV) completion as it cannot cope with some observational effects (dark matter, dark energy, the baryon asymmetry of the universe, etc.), and has a strong sensitivity to the UV scale, which creates a naturalness problem. To solve the latter (a.k.a.~\textit{hierarchy}) problem, different solutions have been introduced, where extra, narrow, resonances are present around, or beyond, TeV energies. However, the elusiveness of experimental data is leading people to find other solutions where there are not any beyond the SM (BSM) narrow resonances. 

The Randall-Sundrum (RS) model~\cite{Randall:1999ee} is a very elegant solution to the hierarchy problem by which, in a five-dimensional (5D) space with a non-factorizable anti de Sitter (AdS) geometry, the Planck and TeV scales are related by the warped factor along the extra dimension. This model has two four-dimensional (4D) boundaries (or hard walls), the UV (or Planck) brane and the infrared (IR) brane (or TeV), and all SM fields are promoted to 5D fields propagating in the bulk of the extra dimension. The theory predicts, on top of the SM fields, the existence of towers of Kaluza-Klein (KK) narrow resonances at the TeV scale.  
However the aforementioned elusiveness of experimental data with new BSM narrow resonances has led to strong bounds on their masses, around 4-5~TeV~\cite{CMS:2018rkg,ATLAS:2019npw}. 

One very interesting possibility, which arises in theories with warped extra dimensions, is that the spectrum of narrow resonances is replaced by a conformal theory with an IR mass gap. Such possibility has been pointed out in Refs.~\cite{Csaki:2018kxb,Megias:2019vdb,Megias:2021mgj,Csaki:2021gfm,Megias:2021arn} for which there is a soft wall in the far IR, which is a naked singularity of the 5D metric, and the discrete KK-spectrum for every 5D field is replaced by an isolated mode (which reproduces the SM field), and a gapped continuum with its same quantum numbers. Moreover it was proven that this configuration corresponds to a particular (critical) 5D metric~\cite{Cabrer:2009we}. In fact, every 5D metric, with a corresponding line element in proper coordinates $y$

\begin{equation}
ds^2=e^{-2A(y)}\eta_{\mu\nu}dx^\mu dx^\nu-dy^2
\end{equation}
is determined by the bulk potential $V(\phi)$, where $\phi$ is a stabilizing 5D (dilatonic) field which is introduced to fix the brane distance and provide a non-vanishing mass to the radion field~\cite{Goldberger:1999un}. In the absence of such stabilizing field, for a constant bulk potential, the metric solution to the 5D Einstein equation is the RS one $A_{\rm RS}(y) = k y$, and the radion is massless, which is phenomenologically excluded. 

In the presence of the field $\phi$ the bulk potential $V(\phi)$ determines, through the gravitational equations of motion, the background metric $A(y)$ and dilaton background profile $\phi(y)$. A common technique, introduced in Ref.~\cite{DeWolfe:1999cp}, to transform the second order equations of motion into a first order system, leads to the introduction of a superpotential $W(\phi)$, related to the bulk potential by 
\begin{equation}
V(\phi)=\frac{1}{8}\left(\frac{\partial W(\phi)}{\partial \phi} \right)^2-\frac{\kappa^2}{6}W^2(\phi).
\end{equation}
The rest of gravitational equations reduce to
\begin{equation}
\phi'(y)=\frac{1}{2}\frac{\partial W(\phi)}{\partial \phi},\quad A'(y)=\frac{\kappa^2}{6}W(\phi)  \,,
\end{equation}
where $\kappa^2=1/(2M_5^3)$, $M_5$ being the 5D Planck scale, and $k$ is a parameter, of the order of the 4D Planck scale $M_{\rm P}$, related to the curvature of the 5D space.

Using the superpotential formalism, the original RS scenario~\cite{Randall:1999ee} corresponds to a constant superpotential $W_{\rm RS}=6k/\kappa^2$, which yields a discrete spectrum of resonances. The model worked out in Refs.~\cite{Megias:2019vdb,Megias:2021arn} has a superpotential given by
\begin{equation}
W=\frac{6k}{\kappa^2}\left( 1+e^{\kappa\phi/\sqrt{3}} \right)  \,,
\end{equation}
where the critical value of the exponent provides a spectrum with a gapped continuum. This theory looks like AdS near the UV brane, while there is a strong departure from conformal invariance near the IR brane. The solution of the gravitational equations of motion yields the background solution 
\begin{equation}
\phi(y)=-\frac{\sqrt{3}}{\kappa}\log[k(y_s-y)],\quad A(y)=ky-\log\left(1-\frac{y}{y_s} \right)  \,,
\end{equation}
where brane potentials $\lambda_\alpha(\phi)$ (with the UV brane corresponding to $\alpha=0$ and the IR brane to $\alpha=1$) fix the values of the field $\phi$ at the branes, $\phi(y_\alpha)=v_\alpha$, $y_s$ is the location of the metric singularity, given by $ky_s=e^{-\kappa v_0/\sqrt{3}}$, while $ky_1=ky_s-e^{-\kappa v_1/\sqrt{3}}$ is the location of the IR brane, where the Higgs is localized, and $y_0 = 0$ is the location of the UV brane. Finally we can identify the scale $\rho$ that controls the mass gap for continuous spectra with the warp factor on the IR brane, \textit{i.e.}~$\rho = k \, e^{-A(y_1)}$, provided that $k(y_s-y_1)=1$, \textit{i.e.}~$v_0=0$~\cite{Megias:2019vdb}. 

In this paper we will contrast the model with the gapped continuum, from Refs.~\cite{Megias:2019vdb,Megias:2021arn}, with experimental data. In particular the gauge boson spectrum contains isolated states, with the mass of the corresponding SM gauge boson, and a gapped continuum above the mass gap $m_g = \rho/2$. Spectra with a similar pattern are obtained for other fields within the present model. The different mass gaps for the different fields are those summarized in Table~\ref{tab:mass_gap}. 
\begin{table}[htb]
\begin{center}
\begin{tabular}{| c | c | c | c | c | c |}
 \hline
  Field & Gauge boson  & Fermion $f$ & Graviton & Radion & Higgs \\ 
 \hline
 $m_g$ & $\rho/2$ &  $|c_f| \rho$ & $3 \rho / 2$ & $3 \rho / 2$ &  $3 \rho / 2$  \\
 \hline
\end{tabular}
\captionof{table}{Values of the mass gap for different fields where $\rho \equiv k \, e^{-A(y_1)}$. The parameter $c_f$ depends on the mass of the fermion (see main text).}\label{tab:mass_gap}
\end{center}
\end{table}

As explained above, the 5D space is contained between the UV brane and the soft-wall singularity. However we have introduced an intermediate brane, the Higgs brane, where the Higgs is living and located such that the hierarchy problem can be solved. While light fermions are localized toward the UV brane, and we can roughly assume they are living on the UV brane, the heavy fermions, in particular the top quark, will be assumed to be localized on the Higgs brane. In this way, for Drell-Yan processes, when light fermions from the protons collide with an EW gauge boson, we can assume that these valence fermions are localized on the UV brane. As for the produced fermions, as they are light, we will also assume that they are localized on the UV brane. The parameter $c_f$ appearing in Table~\ref{tab:mass_gap} controls the localization of the fermion in the extra dimension: $c_f > 1/2$ for light fermions living on the UV brane, and $c_f < 1/2$ for heavy fermions living on the IR brane. Then the mass gap for light fermions turns out to be $m_g > \rho/2$, so that the simplest case for producing the continuum of KK modes are gauge bosons.

Even if there are no resonances, the contribution from the gapped continuum can be seen in an excess  in the value of different cross-sections, which translates into an excess of events, from where one can obtain bounds on the value of the mass gap, which has a common value for all states. In particular in this work we will contrast the model with experimental results from Drell-Yan (DY) processes with dileptons, or one lepton and missing energy, corresponding to the presence of its corresponding neutrino.  As the first step is to contrast predicted cross-sections with corresponding experimental data, we will do in this paper such an exercise with leptons in the final states. We will consider the continuum of electroweak gauge bosons, in particular photon, $Z$ and $W^\pm$. For the photon and $Z$ we will consider as final states dileptons $\ell^+ \ell^-$, while for the $W$ the final states are charged leptons $\ell^\pm\nu_\ell$. 
 
 The outline of this paper is as follows. In Sec.~\ref{sec:model} we present the explicit expressions for the relevant 5D Green's functions which will be used in the numerical analysis of experimental data. In Sec.~\ref{sec:analysis} the numerical analysis is presented, which leads to 95\% bounds on the parameter $\rho$. Finally a summary of results is presented in Sec.~\ref{sec:summary}. Some technical details of the analysis are relegated to the Appendix~\ref{sec:appendix}.

\section{The model predictions}
\label{sec:model}

The Drell-Yan processes for $pp \rightarrow \ell^+ \ell^-$ and $pp \rightarrow \ell^\pm \nu$ are well known and well understood.  This makes them ideal processes to search for an excess (or lack) in dilepton, or charged lepton, final states that do not correspond to a sharply peaked resonance but, instead, a wide one related to the gapped continuous spectrum~\cite{Megias:2021arn}. 

In the SM, we have the various partonic amplitudes:
\begin{align}
    \mM_\gamma &= \bar{u}_3\prth{ie\gamma_\mu} v_4\, P^{\mu\nu}_{A}\,\bar{v}_2 \prth{iQ_q e\gamma_\nu} u_1 \,, \\
    \mM_Z &= \bar{u}_3\prth{ig_\ell\gamma_\mu} v_4\, P^{\mu\nu}_{A,M_Z}\, \bar{v}_2 \prth{ig_q\gamma_\nu} u_1  \,, \\
    \mM_W &= \bar{u}_3 \prth{ig_W \gamma_\mu} v_4\, P^{\mu\nu}_{A,M_W}\, \bar{v}_2 \prth{i g_W V_{q\bar{q}}\gamma_\nu} u_1  \,,
\end{align}
where $u_i$ is a spinor with spin $s_i$ and momentum $p_i$, $V_{q\bar{q}}$ is the corresponding CKM matrix element, and we define the SM propagators for massless and massive bosons, in the Feynman gauge, as
\begin{align}
    P_A^{\mn} &= \frac{\eta^{\mn}}{p^2} \,, \\
    P_{A,M}^{\mn} &= \frac{\eta^{\mn}}{p^2-M^2} \,, 
\end{align}
where $M=(M_Z,M_W)$ is the mass of the corresponding massive gauge boson.

These amplitudes are all easily promoted to 5D, simply by modifying the SM propagators.  
We use the Green's functions \cite{Megias:2021arn}, where $k\lesssim M_{\rm Pl}$ is the curvature of the extra dimension. As we are assuming light fermions both as initial and final states, we can use the UV-to-UV Green's functions, which are given, for LHC momenta and energies ($\ll k$), by
\begin{align}
    G_{A}\prth{y_0,y_0;p}&\stackrel{p \ll k}{\simeq}  -\frac{2k}{\pi p^2} \cdot \frac{J_+\prth{p}}{\Phi\prth{p}} \,, \\
    G_{A,M}\prth{y_0,y_0;p}&\stackrel{p \ll k}{\simeq} -\frac{2k}{\pi p^2} \cdot \frac{J_{M+}\prth{p}}{\Phi_M\prth{p}} \,, 
\end{align}
where $p\equiv \sqrt{p^2}$, and
\begin{align}
   & \Phi\prth{p} = Y_0\prth{p/k} \cdot J_+\prth{p/\rho} - J_0\prth{p/k} \cdot Y_+\prth{p/\rho}  \,, \\
   & \Phi_M\prth{p} = Y_0\prth{p/k} \cdot J_{M+}\prth{p/\rho} - J_0\prth{p/k} \cdot Y_{M+}\prth{p/\rho}  \,, \\
   & J_{\pm} \prth{p/\rho} = 2\frac{p}{\rho} J_0\prth{p/\rho} + \Delta_A^{\pm}J_1\prth{p/\rho}  \,, \\
   & J_{M\pm} \prth{p/\rho} = 2 \frac{p}{\rho} J_0 \prth{p/\rho} + \Xi_A^{\pm}J_1 \prth{p/\rho} \,, \\
   & Y_{\pm}\prth{p/\rho} = 2 \frac{p}{\rho} Y_0\prth{p/\rho}+\Delta^\pm_{A}Y_1\prth{p/\rho}  \,, \\
   & Y_{M\pm}\prth{p/\rho} = 2 \frac{p}{\rho} Y_0\prth{p/\rho} + \Xi_A^{\pm}Y_1 \prth{p/\rho}  \,,
\end{align}
with $J_i(p/\rho)$ and $Y_i(p/\rho)$ being Bessel functions of the first and second kind respectively.
In these expressions we have used the notation
\begin{align}
    \delta_A &= \sqrt{1 - 4p^2 / \rho^2}  \,, \\
    \Delta_A^{\pm} &= \pm\delta_A - 1 \,, \\
    \Xi_A^\pm &= \Delta_A^\pm + 2 ky_s \cdot \prth{m_A/\rho}^2 \,, \\
    ky_s &= ky_1 + 1 \,,
\end{align}
where $ky_s\simeq 36$~\cite{Megias:2021arn} to solve the hierarchy problem.  In the case that $p^2 > (\rho/2)^2$, 
\begin{align*}
\delta_A = -i \sqrt{4p^2/\rho^2 - 1}  \,.
\end{align*}

Multiplying the different $G(y_0,y_0;p)$ propagators by $y_s$ (which transforms the 5D gauge couplings into 4D ones, \textit{e.g.}~the couplings $e,\, g_q,\,g_\ell,\,g_W$), and by the 4D metric $\eta_{\mn}$, we find the propagators we would use for a massless and massive gauge boson respectively. They can be written in terms of the SM propagators, leading to:
\begin{align}
    P_{A5}^{\mn} &= -\frac{2 ky_s}{\pi} \frac{J_+\prth{p}}{\Phi\prth{p}} P_A^{\mn}  \,,  \label{eq:propsA5}\\
    P_{A,M5}^{\mn} &= -\frac{2ky_s}{\pi} \frac{J_{M+}\prth{p}}{\Phi_{M}\prth{p}} \frac{p^2-M^2}{p^2} P_{A,M}^{\mn}  \,, 
\label{eq:props}
\end{align}
thus leading to the 5D amplitudes,
\begin{align}
    \mM_{5\gamma} &= \bar{u}_3\prth{ie\gamma_\mu} v_4 P_{A5}^{\mn} \bar{v}_2 \prth{iQ_q e\gamma_\nu} u_1  \,, \\
    \mM_{5Z} &= \bar{u}_3\prth{ig_\ell\gamma_\mu} v_4 P_{A,M5}^{\mn} \bar{v}_2 \prth{ig_q\gamma_\nu} u_1  \,, \\
    \mM_{5W} &= \bar{u}_3 \prth{ig_W \gamma_\mu} v_4 P_{A,M5}^{\mn} \bar{v}_2 \prth{i g_W V_{q\bar{q}}\gamma_\nu} u_1  \,.
\label{eq:ME}
\end{align}

\section{Analysis}
\label{sec:analysis}

Collider studies in BSM scenarios are usually carried out by encoding the properties and interactions of all new (BSM) particles in MadGraph \cite{MadGraph} (or equivalent), using then the corresponding infrastructure to create Monte Carlo events, apply experimental cuts, and compare with data. Complexities like particle decay, parton showering, and detector effects are applied automatically in a streamlined manner. This path is not possible with the current model, due to the continuum nature of the extra dimension, so a different strategy is needed.

To generate events, we used reweighting. Specifically, we generated SM dilepton and monolepton events via MadGraph, then scaled the weight for each event by the ratio
\begin{align}
	\label{eq:reweight}
    w_i = \frac{\abs{\mathcal{M}}^2_{5D,i}}{\abs{\mathcal{M}}^2_{\SM,i}} w_{\SM,i} \,,
\end{align}
where $\abs{\mathcal{M}}^2_{5D,i}$ and $\abs{\mathcal{M}}^2_{\SM,i}$ are the numerical values of the squared matrix elements in the 5D model and SM for event $i$, obtained by plugging in the four vectors from MadGraph into analytic expressions derived from Eqs.~\eqref{eq:propsA5}-\eqref{eq:ME}, and $w_{\SM,i}$ is the Standard Model weight. The reweight factor $w_i$ is a function of the scale $\rho$, the only free parameter in the 5D theory.

The SM simulated events are then fed through Pythia8 \cite{pythia}, and Delphes \cite{delphes}. Neither of these changes the relative weight, so we can examine distributions of detector level events by rescaling (event by event) the resulting distributions. The only complication is that the rescaling is determined using parton-level events, while the distributions we would like to compare to experiment are formed from detector-level objects, but it is possible to map between these using the event record in Delphes. 

\begin{figure}[h!]
 \centering
 \includegraphics[width=.49\textwidth]{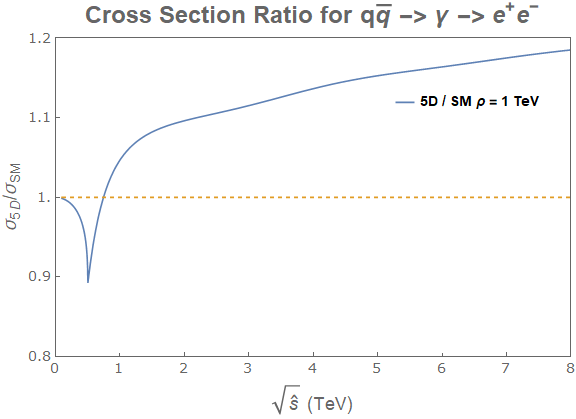}
  \includegraphics[width=.49\textwidth]{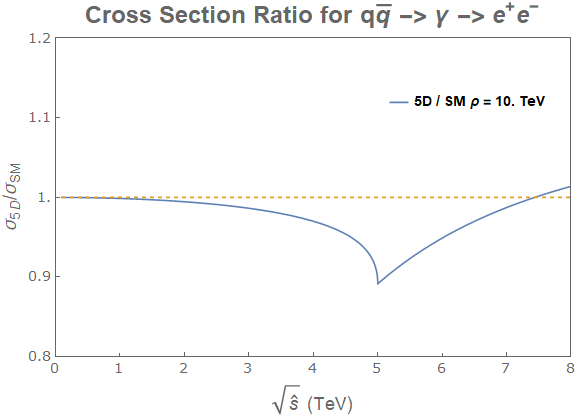}
\caption{Left panel: ratio of the 5D cross section to the SM cross section, with a value of $\rho$ = 1 TeV. Right panel: same ratio, but with $\rho$ = 10 TeV. The $5D/SM$ ratio is less than one for small $\sqrt{\hat s}$, reaching a dip at $\sqrt{\hat s} \sim \rho/2 \equiv m_g$. For larger $\sqrt{\hat s}$, the ratio increases, eventually passing one. The $\sqrt{\hat s}$ range over which $\sigma_{5D}/\sigma_{SM}  < 1$ increases as $\rho$ increases.}
\label{fig:ratio_plot}
\end{figure}

To illustrate the effect of the extra dimension on dilepton and monolepton processes, in Fig.~\ref{fig:ratio_plot} below we display the ratio of partonic cross section, as a function of the partonic center of mass energy $\sqrt{\hat s}$, for $q\bar q \to e^+e^-$ for two different $\rho$ values. For values less (greater) than one, this implies that the 5D model is under (over) predicting the cross section in comparison to the Standard Model.  

\noindent Of particular interest is the feature at the value of $\sqrt{\hat s} = \rho / 2$ which corresponds to the mass gap $m_g$ for gauge bosons~\cite{Megias:2021arn}.  This is related to the structure of the spectral function of the boson.  For $\sqrt{\hat s} < m_g$, we have that the spectral function exactly follows the SM (\textit{i.e.}~a Dirac delta function at the mass of the boson, and zero otherwise). This explains why $\left.\sigma_{5D}/\sigma_{SM}\right|_{\hat s\to 0}=1$ in Fig.~\ref{fig:ratio_plot} which corresponds to the exchange of a (massless) photon. Still for $0<\sqrt{\hat s}<m_g$, even if the imaginary parts of the 5D and SM Green's functions are equal, the real parts are different, as can be easily seen from the explicit expression of the propagators, which explains the difference between $\sigma_{5D}$ and $\sigma_{SM}$ below the mass gap. Above the mass gap though, the continuum modes of the boson change the structure of the spectrum and are thus responsible for the deviation seen in the 5D cross section~\cite{Megias:2021arn}, which explains the dip in the structure of $\sigma_{5D}/\sigma_{SM}$.

For our LHC study, we explored $pp \to e^+e^-, \mu^+\mu^-, e^\pm \nu_e, \mu^\pm \nu_\mu$ at $\sqrt{s} = 13$ TeV with $\mathcal{L} \sim 140 \, \text{fb}^{-1}$ following the experimental analyses~\cite{dileptonbin,chargedleptonbin} as closely as possible. For each process, we determined the reweight factor for $\rho$ values ranging from 1-10 TeV in steps of 0.1 TeV. We generated events using cuts of $m_{\ell \ell} > 200$ GeV and $m_{T, \ell\nu}$ (the transverse mass) $> 100$ GeV to avoid massive boson resonances, and then formed binned distributions of $m_{\ell\ell}$ for the dilepton processes and $m_{T,\ell\nu}$ for the monolepton. For the dielectron (dimuon) channels, we required both leptons to have $p_T \ge 35\, (53)$ GeV and $\abs{\eta} < 2.5\, (2.4)$, while for the monoelectron (monomuon) channels we required lepton $p_T \ge 65\, (55)$ GeV, $\abs{\eta} < 2.47\, (2.5)$, and $\slashed E_T > 50\, \text{GeV}$.  Post cuts, we filtered the Delphes events such that there were exactly 2 (1) outgoing leptons for the dilepton (monolepton) process. For the SM, each surviving event contributes the same weight to the distribution, while for the 5D theory we weight each event by $w_i(\rho)$ in Eq.~(\ref{eq:reweight}).  

To appropriately use the background from the data in \cite{dileptonbin, chargedleptonbin}, we scaled our MadGraph SM histograms to the quoted background values in each $m_{\ell\ell}$, $m_{T,\ell\nu}$ bin. This rescaling accounts for higher order corrections -- present in the SM calculations in  \cite{dileptonbin, chargedleptonbin} but not taken into account by our method -- and differences in the detector response (Delphes versus more realistic, full simulation). We apply the same correction to the 5D events.

 For each value of $\rho$, we calculated a $\Delta \chi^2$ for the 5D model, using observed counts for the dilepton channels from \cite{dileptonbin} and the monolepton channels from \cite{chargedleptonbin}. 
\begin{align}
	\label{eq:deltachi}
	\Delta \chi^2 = \chi^2_{5D} - \chi^2_{\SM},\quad\quad \chi^2 &= \sum_{\bins,i} \frac{(O_i-E_i)^2}{E_i+\sigma_i^2}  \,,
\end{align}
where $O_i$ is the observed data in a given bin, $E_i$ is the value predicted by the model (either SM or 5D model), and $\sigma_i$ is the systematic uncertainty in that bin, taken directly from the experimental analyses. We then use the value of $\Delta \chi^2$ to calculate a bound on $\rho$ to the 95\% confidence level based on the critical value of $\chi^2$ relevant for a single degree of freedom corresponding to $\rho$. We calculate the bound in three different ways, acknowledging the fact that the $\chi^2$ becomes unreliable when the number of events within a bin is small,  using the original binning scheme, a binning scheme in which bins with $<$ 5 events are combined, and a binning scheme where bins with $<$ 10 events are combined.  The calculated bound is set using Eq.~(\ref{eq:deltachi}) with $O \equiv$ LHC observations and $E \equiv$ model predictions.  The expected bound is set using $O \equiv E_{SM}$, the SM prediction, such that $\chi^2_{\SM} = 0$.  Combining all four channels, we find the following limits:

\begin{center}
\begin{tabular}{| c | c | c | c |}
 \hline
  All Channels $\Delta \chi^2$ & Observed Bound & Expected Bound \\ 
 \hline
 Original Binning & 8.5 TeV & 4.2 TeV \\  
 \hline
 $<$ 5 Events Combined & 6.4 TeV & 4.2 TeV \\
 \hline
 $<$ 10 Events Combined & 6.2 TeV & 4.2 TeV \\
 \hline
\end{tabular}
\captionof{table}{Bounds on $\rho$ in TeV from combining all channels. }\label{all_channel_table}
\end{center}
The limits from the individual channels can be found in Appendix~\ref{sec:appendix}.

The combined limit is driven by the $e^+e^-$ channel. In this channel, the SM is not a perfect fit, $\chi^2_{\SM} = 80.03$ for 95 bins, with fluctuations around $m_{\ell\ell} \sim 800\, \text{GeV}$ and $\sim 2~ \text{TeV}$ contributing the most to $\chi^2_{\SM}$. As a result, when setting limits on the 5D model, the $\Delta \chi^2$ is not a monotonic function of $\rho$, as shown below in Fig.~\ref{fig: dielectron_delta_chi}. 
\begin{figure}[H]
    \centering
    \includegraphics[scale = 0.20]{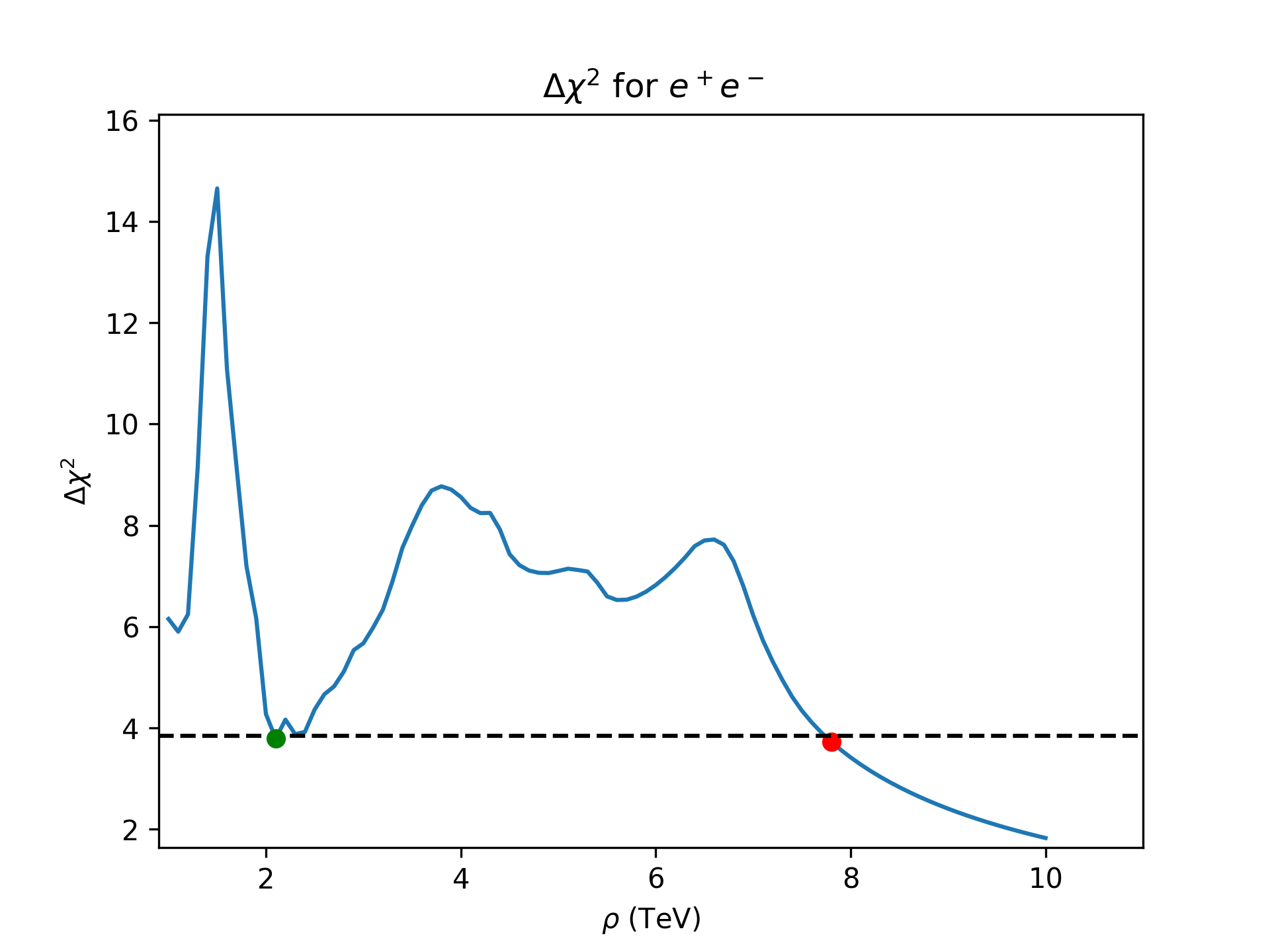}
    \caption{For the process $pp \rightarrow e^+ e^-$, the $\Delta \chi^2$ with the original binning scheme presents two different bounds as seen in Table \ref{dielectron_table}.  In between these two bounds, there is a large region of $\rho$ ruled out to the 95\% level.  The least restrictive bound is seen in green at 2.1 TeV, the most restrictive in red at 7.8 TeV.  The black dashed line is the critical value for 1 dof.}
    \label{fig: dielectron_delta_chi}
\end{figure} 
\noindent For $\rho \sim 2.1, \text{TeV}$ the features of the 5D cross section -- a dip around $\sqrt{\hat s} \sim \rho/2$ followed by a rise -- coincide with the fluctuation in the data to the point that $\Delta \chi^2$ drops below the $95\%$ CL value. For larger or smaller $\rho$, the alignment is spoiled and $\Delta \chi^2$ is bigger. As a result, there are multiple $\rho$ values that satisfy the $95\%$ CL -- $\rho \sim 2.1\, \text{TeV}$ and $\rho \gtrsim 7.8\, \text{TeV}$. We refer to these as the `least restrictive' and `most restrictive' bounds, respectively, and they are listed for each channel individually in Appendix \ref{sec:appendix}\footnote{When there is only one $\rho$ value that satisfies the  $95\%$ CL $\Delta \chi^2$ value, we report the same number for the `least' and `most' restrictive bounds.}. When combining bounds into Table~\ref{all_channel_table}, we only consider the most restrictive bound.

The large difference between the expected and observed bounds is also due to a combination of the relatively poor SM fit and the shape of the 5D cross section relative to the SM. Specifically, in order for the 5D cross section to line up with the SM for $m_{\ell\ell} \sim 800\, \text{GeV}$ (so that data fluctuations in that regime do not generate a large $\Delta \chi^2$) we need $\rho \gtrsim 4\, \text{TeV}$. However, for $\rho$ values of this size, the 5D cross section is suppressed relative to the SM right in the region ($m_{\ell\ell} \sim 2\, \text{TeV}$) where the observed data fluctuates upwards. As a result, the $\Delta \chi^2$ remains high until $\rho$ is large enough that $\sigma_{5D}(m_{\ell\ell} \sim 2\, \text{TeV}) \sim \sigma_{\SM}(m_{\ell\ell} \sim 2\, \text{TeV})$. The bins around $m_{\ell\ell} \sim 2\, \text{TeV}$ have relatively few events, so their impact on the $\Delta \chi^2$ is mitigated somewhat by combining bins, as evidenced by the smaller discrepancy in Table~\ref{all_channel_table} between expected vs.~observed bounds if bins are combined until they contain 5 or 10 events.

In anticipation of the High Luminosity Large Hadron Collider (HL-LHC), we next show the projected bound for $\rho$, assuming that the integrated luminosity ($\mathcal{L}$) of the LHC is increased from $140\, \text{fb}^{-1}$ to $3000\, \text{fb}^{-1}$ -- a factor of $\mathcal L_{\HLLHC}/\mathcal L_{\current} \equiv c = 21$\footnote{To do this rescaling, we are assuming the energy of the HL-LHC is $13\, \text{TeV}$. We do not expect a slight increase in the center of mass energy to affect the bounds significantly.}. For the projected bound, we set $O_i = E_{\SM,i}$ and scale the current (SM) bin counts by $c$. The overall scaling of $\Delta \chi^2$ ($= \chi^2_{5D}$ as the expected $\chi^2_{\SM}$ is zero) depends on how we scale the systematic uncertainties. We explore two different assumptions: i) assuming that the systematic uncertainty percentage $\sqrt{\sigma^2_i}/E_{\SM,i}$ remains constant, and ii) that the systematic uncertainty shrinks with luminosity as $\sigma \to \sigma/\sqrt{c}$. For the first assumption, $\sigma^2_{\HLLHC} = c^2\, \sigma^2_{\current}$, while in the second, $\sigma^2_{\HLLHC} = c\, \sigma^2_{\current}$ -- which leads to the particularly simply scaling $\Delta \chi^2_{\HLLHC} = c\,\Delta \chi^2_{\current}$. We find the following projected bounds:

\begin{center}
\begin{tabular}{| c | c || c | c |}
 \hline
  All Channels $\Delta \chi^2$ & Current Expected Bound & $\sigma^2_{\HLLHC} = c^2 \, \sigma^2_{\current}$ & $\sigma^2_{\HLLHC} = c \, \sigma^2_{\current}$ \\ 
 \hline
 Original Binning & 4.2 & 6.2 & 8.1 \\  
 \hline
 $<$ 5 Events Combined & 4.2 & 6.5 & 8.1 \\
 \hline
 $<$ 10 Events Combined & 4.2 & 6.5 & 8.1 \\
 \hline
\end{tabular}
\captionof{table}{Projected bounds on $\rho$ in TeV from combining all channels with $\mathcal{L} = 3000 \, \textrm{fb}^{-1}$.}\label{all_channel_table_projected}
\end{center}
Thus, extrapolating to the full HL-LHC luminosity raises our expected bound on $\rho$ from 4.2 TeV at the LHC to 6.5 TeV at the HL-LHC (under the more conservative assumption of the luminosity scaling of systematic uncertainties).  This bound is still well within the limits of the HL-LHC, as it is expected to be operating at $\sqrt{s} = 13 - 14$ TeV.

\section{Summary of results}
\label{sec:summary}

Through the application of the 5D model to the Drell-Yan process, we have placed a lower bound on $\rho$ and thus an upper bound on the size of the extra dimension.  This bound was placed using a $\Delta \chi^2$ analysis, reducing the degrees of freedom to 1 corresponding to our parameter $\rho$.  With the current luminosity and data of the LHC, this observed bound is calculated to be 6.2 TeV (4.2 TeV expected), with some dependence on how bins with low numbers of events are handled.  Increasing the luminosity, we project the expected bound to increase to between 6.5 TeV and 8.1 TeV at the HL-LHC, depending on what assumptions are made about the scaling of systematic uncertainties.

\section*{Acknowledgments}
The work of EA is partially supported by the ``Atracci\'on de Talento'' program (Modalidad 1) of the Comunidad de Madrid (Spain) under the grant number 2019-T1/TIC-14019, and by the Spanish Research Agency (Agencia Estatal de Investigaci\'on) through the grants IFT Centro de Excelencia Severo Ochoa No CEX2020-001007-S and PID2021-124704NB-I00, funded by MCIN/AEI/10.13039/\\501100011033. The work of EM is supported by the project PID2020-114767GB-I00 and by the Ram\'on y Cajal Program under Grant RYC-2016-20678 funded by MCIN/AEI/10.13039/501100011\\033 and by "FSE Investing in your future", by the FEDER/Junta de Andaluc\'{\i}a - Consejer\'{\i}a de Economı\'{\i}a y Conocimiento 2014-2020 Operational Programme under Grant A-FQM-178-UGR18, by Junta de Andaluc\'{\i}a under Grant FQM-225, and by the ``Pr\'orrogas de Contratos Ram\'on y Cajal'' Program of the University of Granada. The work of MQ is supported by the Departament d'Empresa i Coneixement, Generalitat de Catalunya, Grant No.~2021 SGR 00649, and by the Ministerio de Econom\'ia y Competitividad, Grant No.~PID2020-115845GB-I00. IFAE is partially funded by the CERCA program of the Generalitat de Catalunya. The work of AD, AM and TS is partially supported by the National Science Foundation under Grant Number PHY-2112540.
The work of RAM has received financial support from CONICET and ANPCyT under projects PICT 2018-03682 and PICT-2021-00374.

\bibliographystyle{unsrt} 
\bibliography{gapped} 

\appendix

\section{Appendix}
\label{sec:appendix}

In this appendix we present the $\Delta \chi^2$ bounds for each channel:  dielectron $pp \to e^+e^-$, dimuon $\mu^+\mu^-$, monoelectron $e^\pm \nu_e$, and monomuon $\mu^\pm \nu_\mu$. We follow the same layout as with the combined bound, presenting the $\Delta \chi^2$ for three different ways of handling bins with low event counts. We also quote both the least and most restrictive bound to cover scenarios, such as in Fig.~\ref{fig: dielectron_delta_chi}, where fluctuations in the observed data can lead to multiple intersections of $\Delta \chi^2$ with the $95\% $ CL line.

\begin{center}
\begin{tabular}{| c | c | c | c |}
 \hline
  Dielectron $\Delta \chi^2$ & Least Restrictive Bound & Most Restrictive Bound & Current Expected Bound \\ 
 \hline
 Original Binning & 2.1 & 7.8 & 2.9 \\  
 \hline
 $<$ 5 Events Combined & 5.1 & 5.1 & 2.9 \\
 \hline
 $<$ 10 Events Combined & 4.9 & 4.9 & 2.9 \\
 \hline
\end{tabular}
\captionof{table}{Bounds on $\rho$ in TeV from the dielectron channel. The $\chi^2_{\SM}$ value for this channel is 80.03 for 95 bins, dropping to 43.38 if bins $<5$ events are combined, and 32.42 if bins with $<10 $ events are combined. }\label{dielectron_table}
\end{center}

\begin{center}
\begin{tabular}{| c | c | c | c |}
 \hline
  Dimuon $\Delta \chi^2$ & Least Restrictive Bound & Most Restrictive Bound & Current Expected Bound \\ 
 \hline
 Original Binning & 2.2 & 2.6 & 3.1 \\  
 \hline
 $<$ 5 Events Combined & 2.2 & 2.6 & 3.1 \\
 \hline
 $<$ 10 Events Combined & 2.3 & 2.7 & 3.1 \\
 \hline
\end{tabular}
\captionof{table}{Bounds on $\rho$ in TeV from the dimuon channel. The $\chi^2_{\SM}$ value for this channel is 14.34 for 42 bins, dropping to 13.12 if bins $<5$ events are combined, and 12.22 if bins with $<10 $ events are combined. }\label{dimuon_table}
\end{center}

\begin{center}
\begin{tabular}{| c | c | c | c |}
 \hline
  Monoelectron $\Delta \chi^2$ & Least Restrictive Bound & Most Restrictive Bound & Current Expected Bound \\ 
 \hline
 Original Binning & 3.0 & 3.0 & 2.9 \\  
 \hline
 $<$ 5 Events Combined & 3.0 & 3.0 & 2.9 \\
 \hline
 $<$ 10 Events Combined & 3.0 & 3.0 & 2.9 \\
 \hline
\end{tabular}
\captionof{table}{Bounds on $\rho$ in TeV from the monoelectron channel.  The $\chi^2_{\SM}$ value for this channel is 29.72 for 49 bins, dropping to 24.5 if bins $<5$ events are combined, and 22.35 if bins with $<10 $ events are combined. }\label{monoelectron_table}
\end{center}

\begin{center}
\begin{tabular}{| c | c | c | c |}
 \hline
  Monomuon $\Delta \chi^2$ & Least Restrictive Bound & Most Restrictive Bound & Current Expected Bound \\ 
 \hline
 Original Binning & 5.3 & 5.3 & 2.2 \\  
 \hline
 $<$ 5 Events Combined & 5.2 & 5.2 & 2.2 \\
 \hline
 $<$ 10 Events Combined & 5.1 & 5.1 & 2.2 \\
 \hline
\end{tabular}
\captionof{table}{Bounds on $\rho$ in TeV from the monomuon channel.  The $\chi^2_{\SM}$ value for this channel is 27.5 for 43 bins, dropping to 23.14 if bins $<5$ events are combined, and 24.31 if bins with $<10 $ events are combined. }\label{monomuon_table}
\end{center}

\noindent Finally, we present the projected bounds for HL-LHC channel by channel. The presentation follows Table~\ref{all_channel_table_projected}, with the first column repeating the expected bound from the current $\mathcal \sim 140\,\text{fb}^{-1}$ dataset.
\begin{center}
\begin{tabular}{| c | c || c | c |}
 \hline
  Dielectron $\Delta \chi^2$ & Current Expected Bound & $c \, \sigma_{\LHC}$ & $\sqrt{c} \, \sigma_{\LHC}$ \\ 
 \hline
 Original Binning & 2.9 & 5.0 & 5.9 \\  
 \hline
 $<$ 5 Events Combined & 2.9 & 5.0 & 5.9 \\
 \hline
 $<$ 10 Events Combined & 2.9 & 5.0 & 5.8 \\
 \hline
\end{tabular}
\captionof{table}{Projected bounds on $\rho$ in TeV from the dielectron channel.}\label{dielectron_table_projected}
\end{center}

\begin{center}
\begin{tabular}{| c | c || c | c |}
 \hline
  Dimuon $\Delta \chi^2$ & Current Expected Bound & $c \, \sigma_{\LHC}$ & $\sqrt{c} \, \sigma_{\LHC}$ \\ 
 \hline
 Original Binning & 3.1 & 5.2 & 6.2 \\  
 \hline
 $<$ 5 Events Combined & 3.1 & 5.2 & 6.2 \\
 \hline
 $<$ 10 Events Combined & 3.1 & 5.2 & 6.2 \\
 \hline
\end{tabular}
\captionof{table}{Projected bounds on $\rho$ in TeV from the dimuon channel.}\label{dimuon_table_projected}
\end{center}

\begin{center}
\begin{tabular}{| c | c || c | c |}
 \hline
  Monoelectron $\Delta \chi^2$ & Current Expected Bound & $c \, \sigma_{\LHC}$ & $\sqrt{c} \, \sigma_{\LHC}$ \\ 
 \hline
 Original Binning & 2.9 & 4.6 & 6.4 \\  
 \hline
 $<$ 5 Events Combined & 2.9 & 4.6 & 6.4 \\
 \hline
 $<$ 10 Events Combined & 2.9 & 4.6 & 6.4 \\
 \hline
\end{tabular}
\captionof{table}{Projected bounds on $\rho$ in TeV from the monoelectron channel.}\label{monoelectron_table_projected}
\end{center}

\begin{center}
\begin{tabular}{| c | c || c | c |}
 \hline
  Monomuon $\Delta \chi^2$ & Current Expected Bound & $c \, \sigma_{\LHC}$ & $\sqrt{c} \, \sigma_{\LHC}$ \\ 
 \hline
 Original Binning & 2.2 & 3.2 & 6.1 \\  
 \hline
 $<$ 5 Events Combined & 2.2 & 3.2 & 6.1 \\
 \hline
 $<$ 10 Events Combined & 2.2 & 3.2 & 6.1 \\
 \hline
\end{tabular}
\captionof{table}{Projected bounds on $\rho$ in TeV from the monomuon channel.}\label{monomuon_table_projected}
\end{center}

\end{document}